# Impact of high-pressure columbite phase of titanium dioxide (TiO$_2$) on catalytic photoconversion of plastic waste and Simultaneous hydrogen (H$_2$) production

Thanh Tam Nguyen[a,b] and Kaveh Edalati[a,b,*]

a  WPI, International for Carbon Neutral Energy Research (WPI-I2CNER), Kyushu University, Fukuoka 819-0395, Japan
b  Mitsui Chemical, Inc. – Carbon Neutral Research Center (MCI-CNRC), Kyushu University, Fukuoka 819-0395, Japan

Photoreforming is a sustainable photocatalytic process that degrades plastic waste while simultaneously producing hydrogen (H$_2$) from water. However, this process has received limited attention due to the scarcity of effective catalysts capable of both plastic degradation and H$_2$ production, such as titanium dioxide (TiO$_2$). In this study, an active catalyst is developed by stabilizing the high-pressure orthorhombic phase of TiO$_2$, known as columbite, using a high-pressure torsion (HPT) method. This material effectively degrades polyethylene terephthalate (PET) plastic under light, converting it into valuable organic compounds such as formic acid, terephthalate, glycolic acid, and acetic acid. Additionally, it produces a significant amount of H$_2$. The findings show that the high-pressure orthorhombic phase, especially in the presence of oxygen vacancies, enhances catalytic H$_2$ production and microplastic degradation by increasing light absorption, reducing electron-hole recombination, and generating hydroxyl radicals. These results highlight the substantial potential of modified high-pressure TiO$_2$ photocatalysts in simultaneously addressing the plastic waste crisis and the demand for H$_2$ fuel.

*Corresponding Author (E-mail: kaveh.edalati@kyudai.jp, Tel: +81-80-3946-4574)

# 1. Introduction

The world is facing various crises and perhaps $CO_2$ emission and plastic waste pollution are the most critical environmental crises. It is intended in many countries, to cut annual global greenhouse gas emissions by using hydrogen ($H_2$) as a clean fuel with no $CO_2$ emissions [1]. Although $H_2$ emerges as a promising energy carrier capable of ushering in a net-zero carbon society, its clean production is a big challenge. $H_2$ can be produced from a variety of resources and technologies; however, the current technologies are either expensive or unsustainable [2]. For example, steam reforming of fossil fuels is currently used to produce 96% of the world's $H_2$ supply, but this process produces large amounts of $CO_2$ [3]. Photocatalytic $H_2$ production using available water and solar energy is a promising approach to producing $H_2$ from renewable energies at room temperature and ambient pressure. However, the method has two main drawbacks: (i) the low efficiency of many catalysts, and (ii) the need for a sacrificial agent for most of the catalysts due to the non-occurrence of oxygen production.

As mentioned above, in addition to the energy and fuel crisis, another environmental issue that the world is facing is the harm that plastic pollution poses to the environment. The release of plastic wastes into the environment leads to their gradual breakdown into small particles, known as microplastics or nanoplastics, raising concerns about their toxic impact on living organisms, including humans [4-6]. To address these energy and environmental problems, the photodegradation of plastic wastes as sacrificial agents for $H_2$ production from the water was introduced as photoreforming by Kawai and his co-workers in 1981 [7].

In the photoreforming process, as schematically shown in Fig. 1a, a catalyst is irradiated by light to excite electrons from the valence band to the conduction band. The electrons in the conduction band contribute to the reduction of water to $H_2$ and the holes in the valence band contribute to the degradation of plastic wastes. In this process, on the one hand, plastic serves as an electron donor and thus no sacrificial agents for $H_2$ production are needed. On the other hand, the oxidation of plastic wastes produces valuable organic chemicals. Despite the environmental benefits of this process, it has received limited attention mainly due to the lack of active catalysts for the process [8-10]. Metal oxides [11-14], bimetallic metal-organic frameworks (MOFs) [15], graphitic carbon nitride [16-18], other polymorphs of carbon nitride [19-24], cadmium sulfide [25-29], and binary alloy sulfides [30-35] are the catalysts used for simultaneous $H_2$ production and plastic degradation. Making composites or using nitrides and sulfides are the main strategies used to enhance the activity of these catalysts for simultaneous $H_2$ production and plastic degradation. However, there are high demands to develop active catalysts from stable and nature-friendly oxides such as titanium dioxide ($TiO_2$), although the application of pure $TiO_2$ for photoreforming is quite limited [7].

Among different catalysts, $TiO_2$ stands out as one of the most common, stable, nature-friendly and effective catalysts for different photochemical applications [36,37]. Different phases of $TiO_2$ such as anatase, rutile and brookite are used for these applications [36,37], while the anatase $TiO_2$ phase with the tetragonal crystal structure is the most favored due to its high surface area and large electron effective mass [38]. However, the activity of anatase for catalysis and particularly for photoreforming is still limited due to its large bandgap of about 3.0-3.2 eV [36-38]. Numerous studies aimed at narrowing the optical bandgap and enhancing the efficiency of $TiO_2$ [38]. These research endeavors have predominantly explored (i) the production of metal-doped oxygen-deficient $TiO_2$ by incorporating elements such as Ag [39], Fe [40], Cr [40], Ni [41], Cu [41], and Ga [41], (ii) the fabrication of nanorods and nanopowders [42], (iii) generation of heterostructures [43], or (iv) the control of crystal structure [44]. While metal doping effectively reduces the bandgap, it may not significantly enhance the activity due to the presence of impurities and related recombination losses [38]. Therefore, in recent years, attention has shifted towards narrowing the optical bandgap of pure $TiO_2$ using dopant-free approaches [44-46]. Such approaches have led to the development of new catalysts exhibiting remarkable catalytic performance for $CO_2$ conversion [47,48] and $H_2$ production [49], but there have been few attempts to use these approaches to enhance photoreforming.

In this study, an active catalyst is synthesized for $H_2$ production and plastic degradation by stabilizing a high-pressure phase of $TiO_2$ and without the addition of impurity atoms. As shown in Fig. 1b, $TiO_2$ has a high-pressure phase with an orthorhombic crystal structure (known as $TiO_2$-II or columbite) [50] which has not been investigated as a catalyst for photoreforming. This phase is stabilized in this study by the high-pressure torsion (HPT) method (Fig. 1c) [51], a severe plastic deformation process that has a high potential to regulate the polymeric phase transformations in various oxides, including ZnO [52], $ZrO_2$ [53], and $TiO_2$ [48]. This first application of the high-pressure $TiO_2$ phase for photoreforming confirms its high activity for both polyethylene terephthalate (PET) plastic degradation and $H_2$ production.

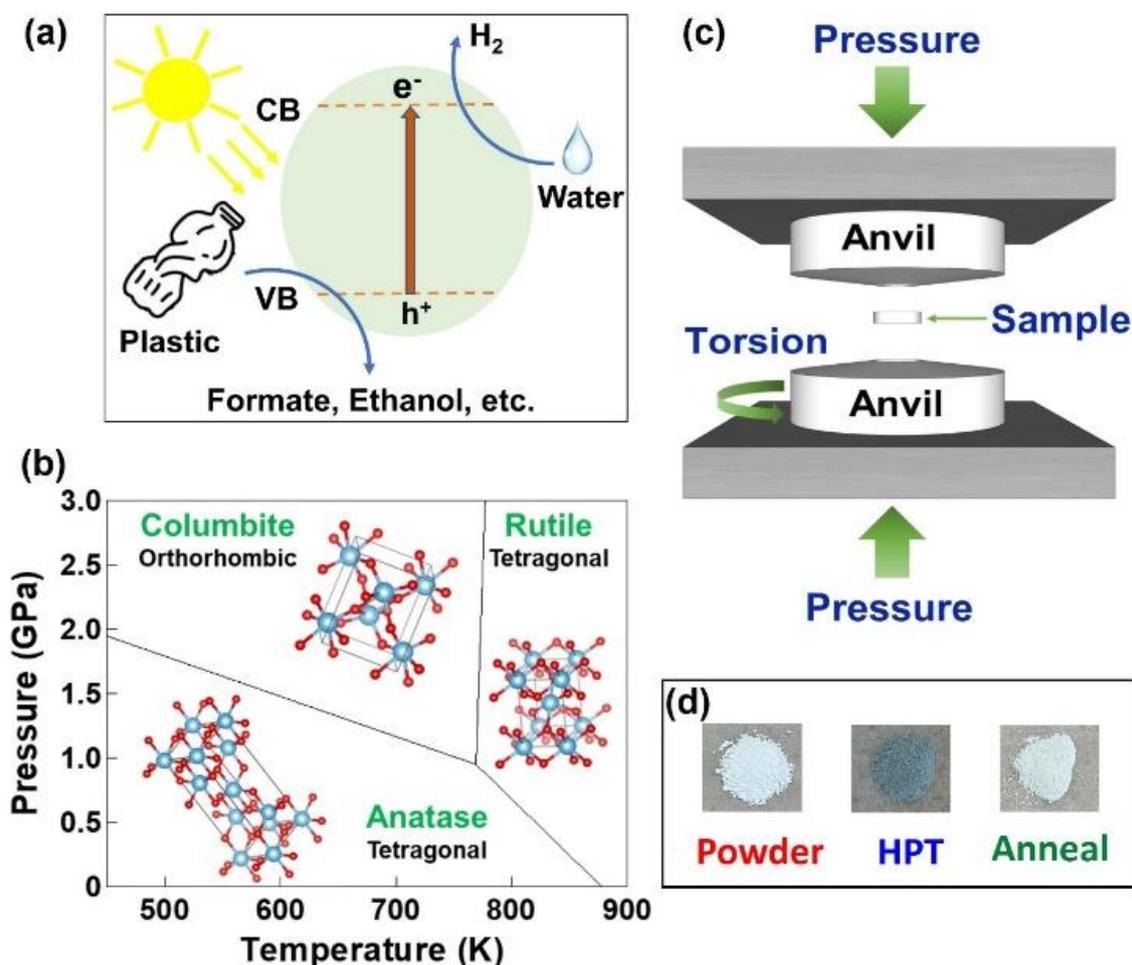

**Figure 1.** (a) Illustration of simultaneous plastic degradation and $H_2$ production by photoreforming, (b) phase diagram of $TiO_2$ under pressure and temperature, and (c) illustration of HPT method.

## 2. Experimental Procedures
### 2.1. Reagents
Anatase with a purity of 99.8% was purchased from Sigma-Aldrich Company, USA. NaOH was purchased from Fujifilm, Japan and a solution of 10 M NaOD (40 wt% in $D_2O$ with a purity of 99 at%) was also obtained from Sigma-Aldrich. Micro PET plastic powder with a particle size of 300 μm was purchased from GoodFellow Company, UK. Coumarin and Maleic acid were obtained from Tokyo Chemical Industry Co., LTD and Nacalai Tesque, Inc., Japan.

## 2.2. Preparation of catalysts

To synthesize the high-pressure phase, 250 mg of anatase powder was compressed under 380 MPa into a 10 mm diameter disc. The compressed disc was then put into the circular hole (10 mm diameter and 0.25 mm depth) of the lower anvil of an HPT machine. The lower anvil was slowly raised to touch the upper anvil until a pressure of 6 GPa was introduced to the sample. The HPT process was performed by rotating the lower anvil against the upper anvil for $N = 3$ turns with a rotation speed of 1 rpm under 6 GPa. Since the HPT method can generate oxygen vacancies [38,39], some amounts of HPT-processed samples were annealed at 773 K for 1 h to change the amount of oxygen vacancies. After these processes, the three samples (initial powder, HPT-processed sample and annealed sample) were manually crushed and examined by different characterization methods and photoreforming tests. The appearance of the three samples is shown in Fig. 1d, indicating that the color of the sample changes to dark after HPT and returns to white after annealing, probably due to the formation of vacancies by HPT processing and their annihilation by annealing.

## 2.3. Catalyst Characterization

The crystalline structures of the three catalysts were characterized by X-ray diffraction (XRD) by Cu Kα irradiation with a wavelength of $\lambda = 0.1542$ nm. The crystallite size of catalysts was calculated by the Halder-Wagner method [54]. Raman analysis was conducted by using a 532 nm laser to investigate the crystal structures. For microstructural analyses, the samples in small quantities were crushed in ethanol and dispersed on carbon grids for transmission electron microscopy (TEM). Microstructural analyses were conducted in the bright-field (BF) mode, dark-field (DF) mode, selected area electron diffraction (SAED), and high-resolution (HR) imaging mode combined with fast Fourier transform (FFT). Light absorbance of the three catalysts was measured by ultraviolet-visible (UV-vis) diffuse reflectance spectroscopy within the range of 200-900 nm. Bandgaps were determined by the Kubelka-Munk analysis. The specific surface area of catalysts was measured by nitrogen gas adsorption using the Brunauer-Emmett-Teller (BET) method. The specific surface areas were 10.20, 0.67 and 1.42 $m^2/g$ for the initial powder, HPT-processed sample and annealed sample, respectively.

X-ray photoelectron spectroscopy (XPS) was performed using an Al Kα source to study the electronic states of oxygen and titanium, determining the generation of oxygen vacancies and estimating the bottom of the valence band. Electron paramagnetic resonance (EPR) was conducted at ambient temperature using a 9.4688 GHz microwave source to analyze the electron spins and assess the presence of oxygen vacancies and the formation of $Ti^{3+}$ states. The recombination of electrons and holes was examined by steady-state photoluminescence (PL) emission spectroscopy with a 325-nm laser source.

## 2.4. Photocatalytic Experiments

For catalytic $H_2$ production and plastic degradation, 50 mg of catalyst and 50 mg of PET powder with the size of 300 μm were mixed with 3 mL solution of 10 M NaOH. Additionally, $Pt(NH_3)_4(NO_3)_2$ was introduced as a source of 1 wt% platinum co-catalyst. The mixture underwent sonication for a duration of 5 minutes and was then purged with argon for 30 minutes. After air evacuation, the mixture was subjected to 300 W Xenon light irradiation while being continuously stirred at a constant temperature of 298 K. The production of $H_2$ was measured by a gas chromatography (GC) facility equipped with a thermal conductivity detector. The photo-oxidized products resulting from the degradation of PET plastic were identified through $^1H$ nuclear magnetic resonance (NMR) spectra using a solution of 10 M NaOD in $D_2O$. To quantitatively analyze the yield of photo-oxidized products by $^1H$ NMR, maleic acid was employed as an internal standard [19]. Furthermore, the extent of produced hydroxyl (•OH) radicals was determined by adding coumarin (1 mM) to the solution and measuring the fluorescence signal intensity at 450 nm corresponding to 7-hydroxycoumarin using a spectrofluorometer with an excitation wavelength of 332 nm [55].

## 3. Results
### 3.1. Catalyst Characterization

The phase transformations of anatase powder under HPT and HPT followed by annealing were confirmed by XRD and Raman spectroscopy. Fig. 2a shows the XRD spectra of the three catalysts. In the initial powder, anatase peaks are clearly observed, but new peaks corresponding to the high-pressure phase appear after HPT processing and remain after annealing. A peak for the (111) plane of the high-pressure phase at $2\theta = 31.4°$ is the most distinguished peak for phase transformation in both HPT and annealed samples. Besides the high-pressure phase, new peaks of rutile also appear under the HPT and annealing conditions. The broadening of peaks also occurs after HPT processing, indicating the formation of lattice strain and defects and the reduction of crystallite size. With annealing, the peak intensity for the high-pressure phase slightly decreases, and the peak broadening becomes less intense. The quantitative analysis of phase fractions, crystallite size and lattice strain for all three samples is summarized in Table 1. The transformation of anatase to a defective high-pressure phase by HPT should be due to the simultaneous effect of high pressure and torsional strain [51-53].

Raman spectra, shown in Fig. 2b, confirm that the initial powder has an anatase phase. After HPT processing at 6 GPa, the high-pressure phase forms and remains stable under ambient pressure even after annealing at 773 K. Additionally, there are other peaks corresponding to the rutile phase after HPT and annealing, which is in line with the XRD results. The stability of the high-pressure phase at ambient pressure and the formation of the rutile phase at a temperature lower than the predicted temperature by the phase diagram suggests that HPT-induced straining plays an important role in $TiO_2$ phase transformations and stabilization [56].

The microstructural and nanostructural features of the three catalysts are shown in the TEM images of Fig. 3 with BF, DF, SAED, and HR imaging combined with FFT. The initial anatase powder has an average grain size of 150 nm. Due to partial consolidation, the sizes of individual particles increase by HPT processing, but the grain sizes reduce to the range of nanometers. The reduction of grain size after HPT processing can be also recognized by the diffraction pattern changes from a dot form to a ring form in SAED images. HR images confirm that there are many nanograins of anatase, columbite, and rutile in samples after HPT processing and annealing which is in good agreement with quantitative XRD analyses in Table 1. The crystal size reduction by HPT should be due to the severe plastic deformation effect induced by the process [57,58].

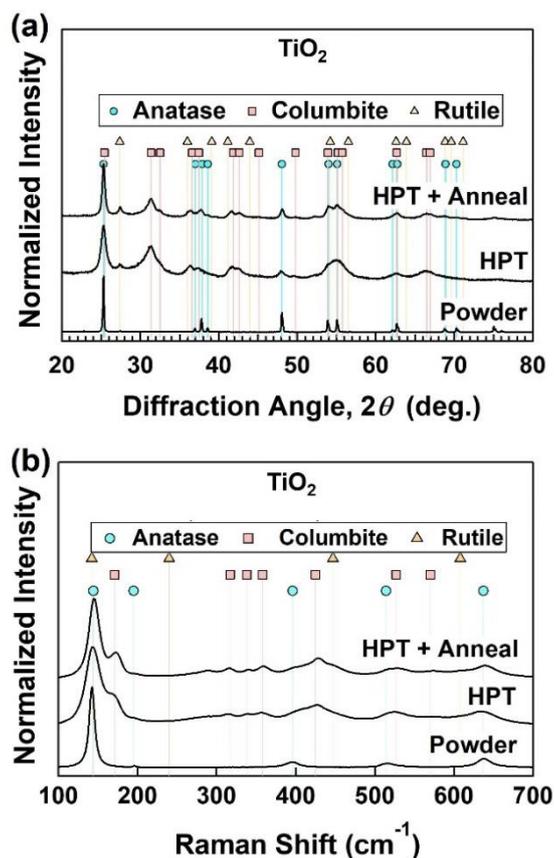

Figure 2. Formation of high-pressure TiO$_2$ phase after HPT processing and its stability during annealing. (a) XRD and (b) Raman spectra of initial powder, HPT-treated, and annealed samples.

Table 1. Phase fractions, average crystallite size, and lattice strain determined from XRD profiles.

| Sample | Phase Fractions (wt%) | Crystallite Size (nm) | Lattice Strain (%) |
|---|---|---|---|
| **Powder** | Anatase (97.5%), Rutile (2.5%) | 55 | 0 |
| **HPT** | Anatase (18.5%), Rutile (13.3%), Columbite (68.2%) | 6 | 0.9 |
| **HPT + Anneal** | Anatase (29.8%), Rutile (14.1%), Columbite (56.1%) | 10 | 0.6 |

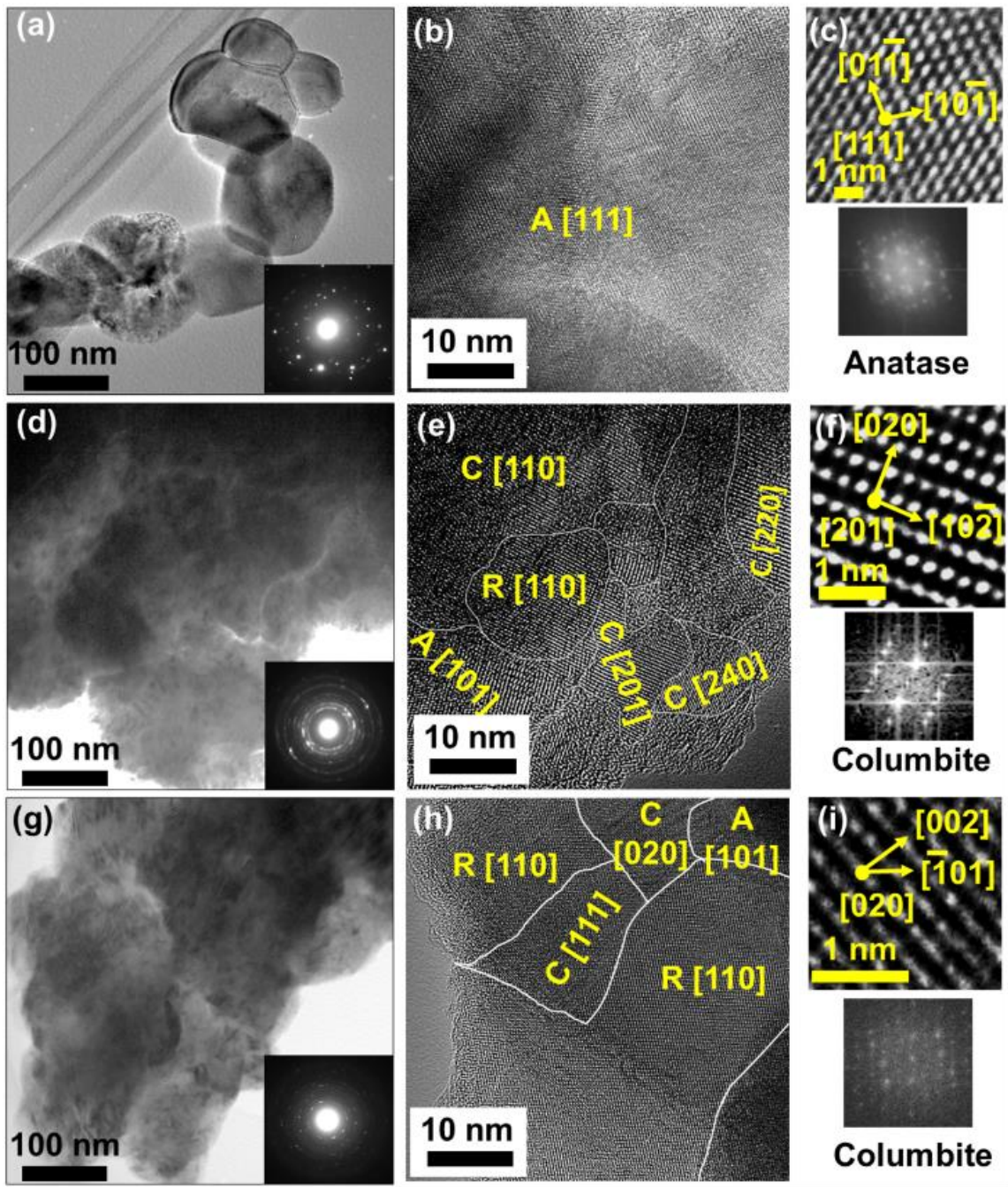

Figure 3. Existence of nanograined high-pressure TiO$_2$ phase after HPT processing and annealing. TEM (a, d, g) BF images and corresponding SAED patterns, (b, e, h) HR images, and (c, f, i) lattice images and corresponding FFT diffractogram for (a-c) initial powder, (d-f) HPT-treated sample and (g-i) annealed sample. A: anatase, C: columbite, and R: rutile.

XPS was used to investigate the formation of oxygen vacancies, as demonstrated in Fig. 4 for (a) O 1s and (b) Ti 2p. By taking into account the C 1s peak location at 248.8 eV, the XPS peak positions were adjusted. After HPT processing, peak positions for titanium and oxygen do not change, but there are noticeable shoulders to lower energies in the Ti 2p XPS spectrum and to higher energies for O 1s XPS spectrum, as shown more clearly in magnified views of spectra in Fig. 5c and 5d. The spectra return to normal conditions, similar to the spectra of the initial anatase powder, after further annealing. The shoulders after HPT processing should be due to the formation of oxygen vacancies and their disappearance after annealing should be due to the annihilation of vacancies. These observations are consistent with the color changes in the samples which is darker after HPT processing due to the formation of oxygen vacancies [59,60].

EPR was used to investigate the nature of oxygen vacancies, as shown in Fig. 5. The characteristic peaks for anatase can be seen in the initial powder [61]. These characteristic peaks become so weak after HPT processing, and several pair peaks that correspond to different point defects appear. For example, the peak with a *g* value of 2.006 is associated with hole-trapping sites, such as oxygen vacancies or oxygen-centered radicals, while the peaks with *g* values smaller than 2 are associated with electron trapping sites, such as $Ti^{3+}$ radicals [61-63]. These pair peaks vanish after annealing, signifying the elimination of oxygen vacancies during heating. It should be noted that while oxygen vacancies can serve as reaction or recombination sites, their presence on the surface of catalysts is usually advantageous for enhancing activity.

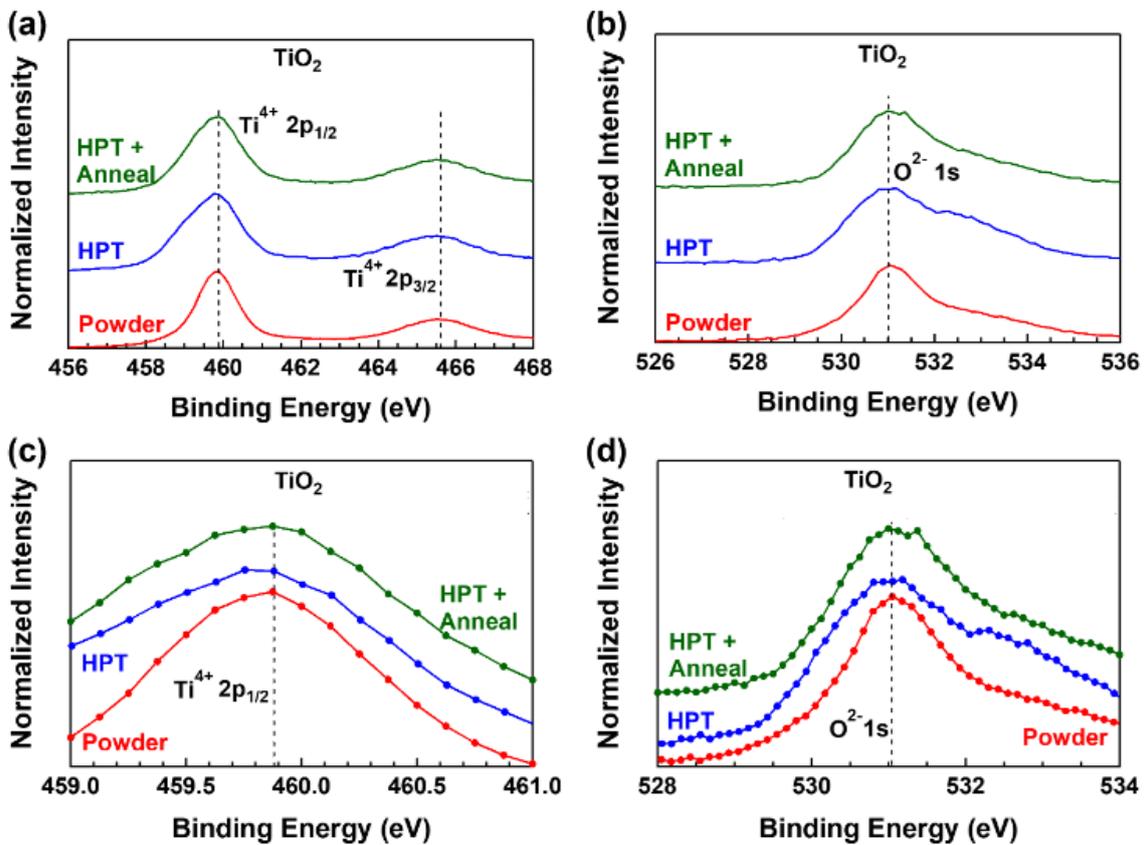

Figure 4. Formation of oxygen vacancies by HPT processing and their disappearance by annealing. XPS spectra with (a, c) Ti 2p (b, d) O 1s for initial powder and HPT-treated and annealed samples, where (c) and (d) are magnified views of (a) and (b) respectively.

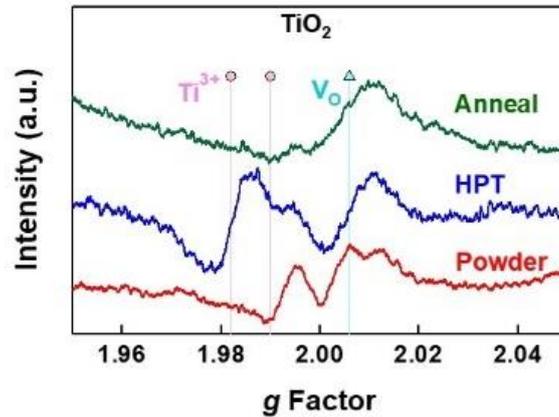

Figure 5. Formation of oxygen vacancies by HPT processing and their disappearance by annealing. EPR spectra for initial powder and HPT-treated and annealed samples.

The UV-vis spectra in Fig. 6a, with the light absorption coefficient plotted against wavelength, reveal that the three examined samples primarily exhibit large UV light absorption characteristics. Within the visible light region, there is a discernible increase in light absorbance for the HPT sample and a subsequent diminishing in light absorbance for the annealed sample. Such tail-like light absorbance in the visible light region usually appears in catalysts with large oxygen vacancy concentrations [46,53,57]. In addition to the enhanced light absorbance in the visible light region, the absorbance edge also shifts to lower energies after HPT processing. This phenomenon is further elucidated in Fig. 6b, where bandgap calculations are performed using the Tauc equation as described by $(\alpha h\nu)^{1/r} = A(h\nu - E_g)$, in which $A$ is a constant, $r$ is a constant depending on the bandgap type (2 for indirect bandgap and 0.5 for direct bandgap) and $h\nu$ is the phonon energy ($\alpha$: light absorption, h: Planck's constant, $\nu$: light frequency) [64]. Since $TiO_2$ possesses an indirect bandgap, an $r$ value of 2 should be used for the Tauc plot [48]. The initial powder possesses a bandgap of 3.19 eV, consistent with the reported bandgap of anatase [36-38]. After HPT processing, the bandgap decreases to 2.77 eV, which should be due to the formation of the high-pressure $TiO_2$ phase decorated with oxygen vacancies. Conversely, the subsequent annealing of the HPT-treated samples leads to an increase in bandgap from 2.77 eV to 2.90 eV which should be due to the annihilation of vacancies [65]. The large light absorbance of the HPT-treated sample is considered a positive feature for $TiO_2$ catalysts [48,49]. The XPS profiles used to determine the top of the valence band are shown in Fig. 6c. The bottom of the conduction band can be determined by subtracting the bandgap from the top of the valence band to achieve the electronic band structure, as shown in Fig 6d. HPT-processed sample shows an increase in the energy level of the top of the valence band together with a bandgap narrowing. The band structure of all three catalysts satisfies the energy requirement for hydrogen, oxygen, and hydroxyl radical production.

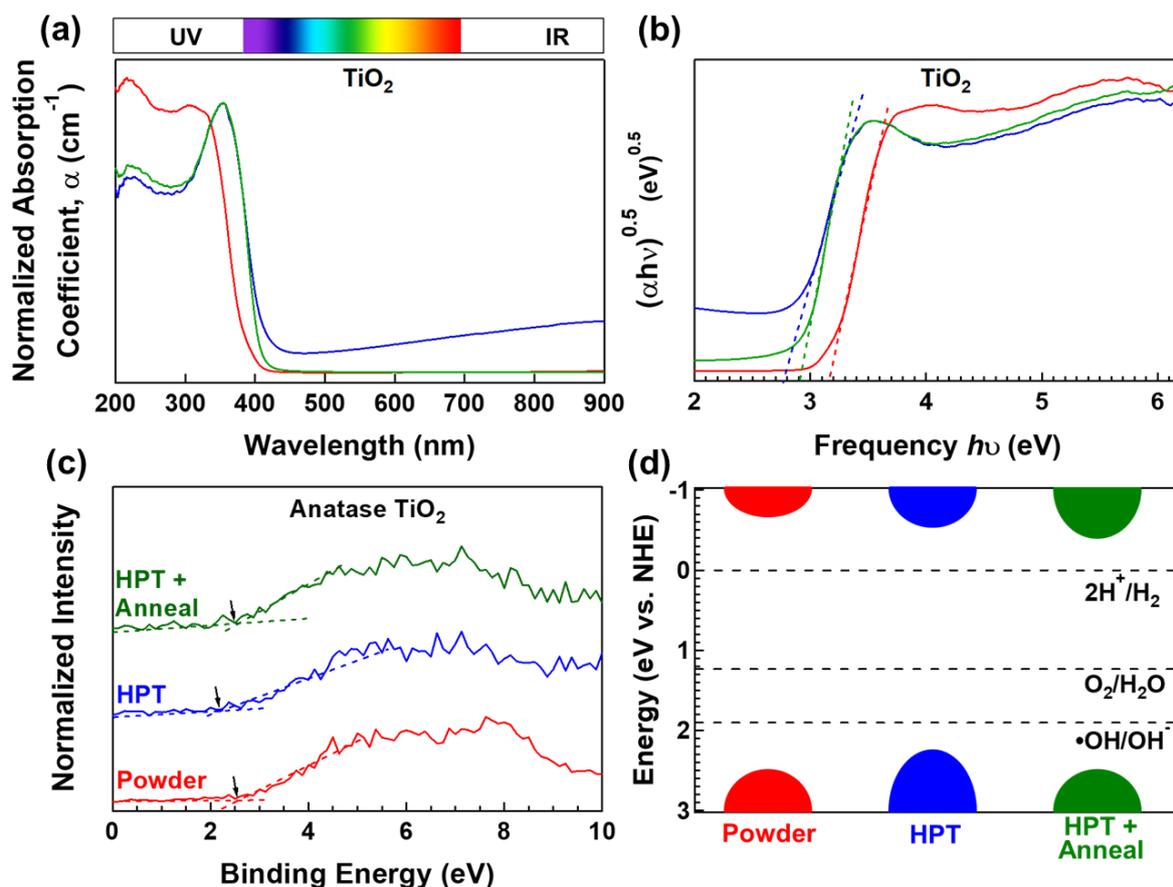

Figure 6. Large light absorbance and narrowing of optical bandgap by introducing high-pressure $TiO_2$ phase. (a) Light absorption coefficient versus light wavelength achieved using UV-vis absorbance spectroscopy, (b) Tauc plot for bandgap calculation ($\alpha$: light absorption, h: Planck's constant, $\nu$: light frequency), (c) XPS of top of the valence band and (d) band structure compared to energy requirements for hydrogen, oxygen and hydroxyl radical formation for initial powder and HPT-treated and annealed samples.

To examine the radiative electron-hole recombinations, the steady-state PL spectra were achieved, as shown in Fig. 7. PL intensities for the HPT-processed and annealed samples are noticeably lower than those of anatase powder. This implies that compared to the initial anatase powder, there is s suppression of electron and hole recombination in these two samples. Such suppression may not be due to oxygen vacancies because the annealed sample does not contain large fractions of oxygen vacancies; however, it should be due to the effect of the high-pressure phase which is high in both samples.

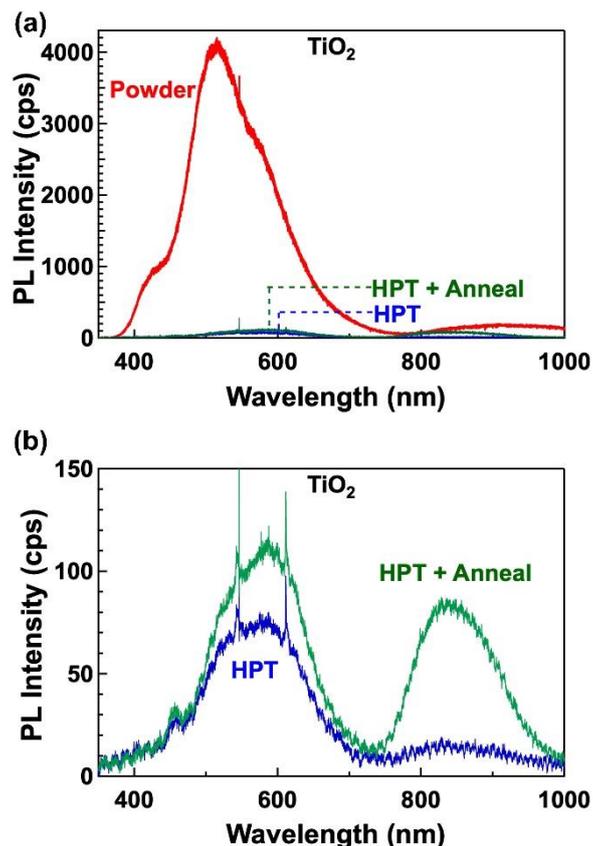

Figure 7. Suppression of radiative electron-hole recombination by high-pressure $TiO_2$ phase formation. PL spectra for (a) initial powder and HPT-treated and annealed samples, and (b) HPT-treated and annealed samples in enlarged view.

### 3.2. Photoreforming

During the photoreforming process in the aqueous solution, both $H_2$ production and PET degradation occur. The catalytic $H_2$ production of the three catalysts under UV irradiation is presented in Fig. 8. As can be seen, the production of $H_2$ does not occur at time zero when the catalyst was stirred in the solution in the dark for 1 hour. Moreover, no $H_2$ is detected in the blank test when no catalysts are added to the solution under UV illumination. After HPT processing, the $H_2$ production amount significantly increases and slightly decreases after annealing at 773 K. The best catalytic activity for $H_2$ production is achieved from the HPT sample with the amount of 1.8 mmol m$^{-2}$ after 4 h which is about 4.5 times better than that of initial anatase powder. Such an improvement is particularly important because it could be achieved by a dopant-free strategy, which is currently of interest [44-49]. It should also be noted that, to understand the effect of the co-catalyst on photoreforming, the photoreforming process using HPT-induced columbite and without platinum addition was also conducted. Data, as shown in Fig. 8, indicate that hydrogen is produced in the first 30 min of irradiation and the production stops in longer irradiation periods. The hydrogen production significantly increases with co-catalyst addition, confirming the important role of platinum co-catalyst for $TiO_2$ in the photoreforming process, similar to other photocatalytic processes. In the photocatalytic system, platinum co-catalyst serves as an electron trap site, reducing the recombination of electrons with holes [37-42]. Moreover, the plutonium co-catalyst provides active sites for the reduction reaction of protons to $H_2$ molecules. The addition of a co-catalyst also lowers the activation energy, subsequently enhancing the overall efficiency of the photocatalytic process [37-42].

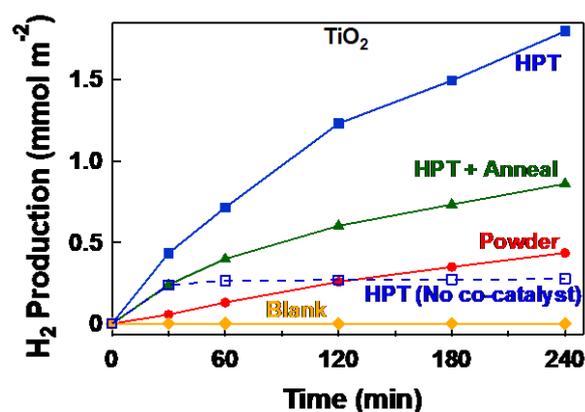

Figure 8. Enhancement of $H_2$ production from photoreforming of PET plastic by the introduction of high-pressure $TiO_2$ phase. $H_2$ production versus irradiation time during catalytic process with PET plastic as sacrificial agent for initial anatase powder and catalysts treated by HPT and annealing with platinum co-catalyst addition. HPT-processed was also examined with no co-catalyst addition.

To identify the degradation products of PET, NMR was employed for the analysis of the liquid phase. Fig. 9 illustrates $^1$H NMR spectra of three catalysts: (b) initial powder, (c) HPT-treated sample, and (d) annealed sample. PET in NaOH conditions undergoes hydrolysis to its monomers of terephthalate and ethylene glycol [18,19,22,25]. Ethylene glycol is further oxidized by the catalyst to form other useful products, as shown in Fig. 9e. The photoreforming process using the starting anatase powder results in the formation of acetic acid, while using catalysts prepared by HPT and annealing generates formic acid, glycolic acid and acetic acid. Quantitative $^1$H NMR data are presented in Fig. 10 to investigate the PET degradation degree. These data reveal that the powder sample produces 1.33 mmol m$^{-2}$ terephthalate and 2.93 mmol m$^{-2}$ acetic acid. The introduction of the high-pressure phase under HPT processing not only increases the $H_2$ production efficiency but also enhances the oxidation process of PET. The amounts of oxidized products obtained from the HPT-treated sample within 4 h are 8.25, 2.36, 9.90, and 24.7 mmol m$^{-2}$ formic acid, terephthalate, glycolic acid and acetic acid, respectively. These degradation rates suggest that at least 17% of PET is degraded within 4 h of irradiation (i.e. full degradation in 24 h). Since some photodegradation products may not be detectable by NMR, the real degree of photodegradation of PET microplastics can be even higher than 17% per 4 h. These results confirm the high potential of photoreforming in the degradation of plastic waste while simultaneously generating green $H_2$, which can be of interest due to the current environmental and energy crises [1,3-6].

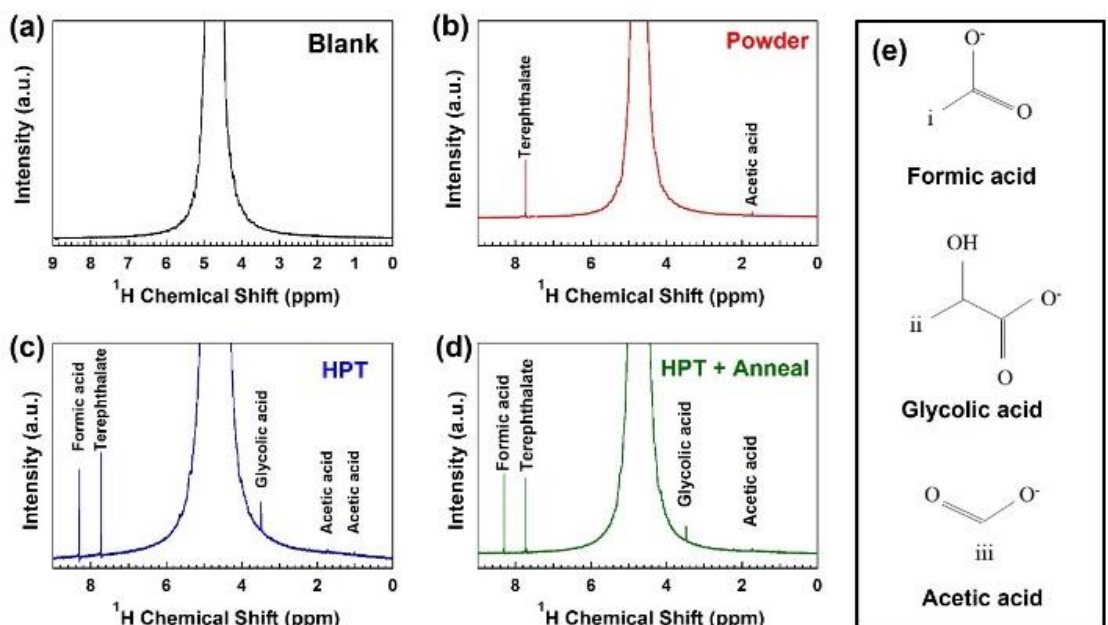

Figure 9. Photoreforming of PET plastic to formic acid, terephthalate, glycolic acid and acetic acid by $TiO_2$ catalysts. $^1$H NMR spectra of degraded PET after 4 h irradiation for (a) blank test, (b) initial powder, (c) HPT-treated catalyst and (d) annealed catalyst in 10 M NaOD in $D_2O$. (e) Chemical formula of the products.

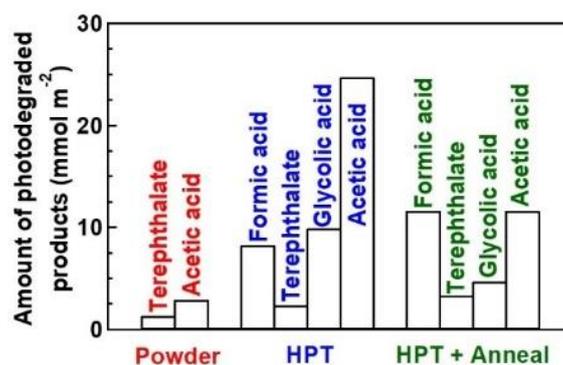

Figure 10. Degradation of PET plastic to small molecule hydrocarbons by photoreforming. Quantitative $^1$H NMR analysis for the concentration of formic acid, terephthalate, glycolic acid, and acetic acid produced by the degradation of PET after 4 h UV irradiation using initial anatase powder and catalysts processed by HPT and annealing in a solution of 10 M NaOD in $D_2O$ with maleic acid as internal standard.

Since photoreforming reactions involve oxidation transformation, it is essential to evaluate the production of the main oxidants which are •OH radicals. The •OH radicals are essential in this process for breaking down PET plastic into smaller organic compounds. To evaluate the •OH radicals, coumarin as a probe radical-trap chemical was added during the photoreforming. As shown below, •OH radicals react with coumarin to form 7-hydroxycoumarin which can be detected by fluorescence spectroscopy [66].

•OH + Coumarin → 7-Hydroxycoumarin      (1)

Fig. 11 provides a comparative analysis, showcasing the differences in the amount of •OH radicals generated by the initial powder and HPT-treated and annealed catalysts. The data reveal a discernible increase in the •OH radical production by HPT treatment, while the •OH radical production decreases after annealing. These observations suggest that the HPT-treated catalysts should show the highest catalytic activity and the initial powder should show the lowest activity for oxidation of PET due to the presence of larger amounts of reactive radicals [67].

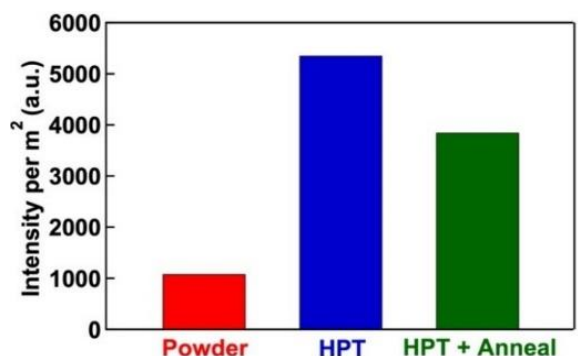

Figure 11. Enhanced formation of •OH radicals during photoreforming by the introduction of high-pressure $TiO_2$ phase. Fluorescence signal intensity at 450 nm corresponding to •OH radical concentration produced during photoreforming after 20 minutes of irradiation using initial anatase powder and catalysts processed by HPT and annealing.

In order to investigate the stability of the catalysts during catalytic $H_2$ production and PET plastic degradation, an assessment was conducted by Raman spectroscopy before and after photoreforming. As depicted in Fig. 12, the crystal structures of all examined catalysts remain stable through the processes, and no discernible alterations are detected in the Raman spectra. The high stability of the high-pressure phase for simultaneous $H_2$ production and plastic degradation expands the application of this less-known phase of $TiO_2$ as an active catalyst [58,60,67].

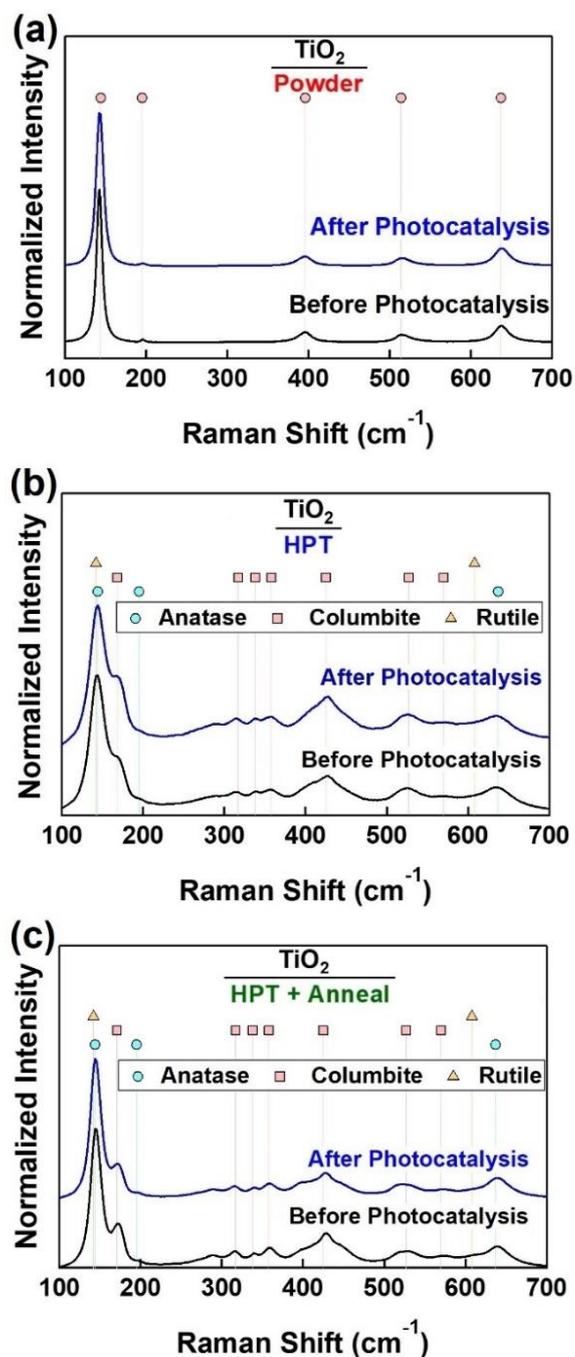

Figure 12. Stability of examined catalysts for photoreforming and plastic degradation. Raman spectra of (a) initial powder, (b) HPT-treated catalyst and (c) annealed catalyst before and after photoreforming test for 4 h irradiation.

## 4. Discussion

The results presented in this study show the high activity of the high-pressure $TiO_2$ columbite phase for photoreforming and simultaneous plastic degradation and $H_2$ production. Two questions need to be discussed further concerning this finding. First, what are the mechanisms of simultaneous $H_2$ production and PET plastic degradation? Second, how is the activity of the high-pressure phase containing $TiO_2$ compared with the best catalysts in the literature?

Regarding the first question, it should be noted that the $H_2$ production from the photoreforming of plastic waste is similar to the conventional photocatalytic $H_2$ production process using sacrificial agents. In this process, PET plastic acts as a sacrificial electron donor, oxidized by holes from the valence band (VB) or reactive radicals. Simultaneously, water is reduced to generate H$_2$ by photoexcited electrons in the conduction band (CB). The sequence of events is elucidated through the series of reactions as follows [2,68].

$$Semiconductor + Light \rightarrow h^+_{(VB)} + e^-_{(CB)} \tag{2}$$

$$h^+_{(VB)} + H_2O \rightarrow \bullet OH + H^+ \tag{3}$$

$$2H^+ + 2e^- \rightarrow H_2 \tag{4}$$

$$h^+_{(VB)} / \bullet OH + Plastics \rightarrow Organic\ Products \tag{5}$$

In the structure of the HPT-processed sample, the high-pressure phase of TiO$_2$ narrows the optical bandgap [48,49]. For the annealed sample, removing oxygen vacancies leads to a decrease in catalytic activity compared to the HPT-processed sample, suggesting that the presence of oxygen vacancies in the high-pressure phase is of significance in enhancing the photoreforming activity. The oxygen vacancies can usually serve as shallow charge carrier traps, extending the lifetime of electron-hole pairs and broadening the range of phonon absorption [69]. The excess electrons generated from oxygen removal are usually transferred to the empty 3d orbitals of metals, forming shallow donor states below the conduction band, thereby enhancing light absorption [70]. Moreover, oxygen vacancies can act as active sites for catalytic reactions [69]. In an earlier theoretical study [71], the first-principles calculations using density functional theory (DFT) suggested that the bandgap of the columbite phase is not narrower than that of other TiO$_2$ polymorphs (3.4 eV for columbite versus 3.2 eV for anatase). However, columbite exhibits a lower energy of vacancy formation (3.1 eV for columbite versus 5.8 eV for anatase), allowing oxygen vacancies to form more easily in this high-pressure phase. DFT calculations also showed that the optical bandgap narrowing (i.e. light absorbance) of the columbite phase by oxygen vacancy generation is more significant than that of the anatase phase. Another issue reported by DFT calculations is the high surface activity of the columbite phase. For example, the (101) surface of columbite, analyzed using the nudged elastic band and climbing image nudged elastic band methods, demonstrates a strong affinity for water molecule adsorption and a low activation energy for splitting water into hydrogen and hydroxyl radical groups. Such enhanced hydrogen and hydroxyl radical formation on the oxygen-deficient columbite phase contribute to hydrogen production and plastic degradation, respectively [2,68]. Removal of oxygen from the crystal leads to a broken bond, known as a dangling bond, which creates energy states in the band structure, leading to a higher light absorbance [72,73]. The formation of the hydroxyl group also changes the optical bandgap [74]. It should be noted that when impurity atoms generate oxygen vacancies and dangling bonds, they may lead to undesirable effects, such as impurity-induced charge carrier recombination, which diminishes these benefits [70]. Importantly, in this study, the oxygen vacancies in columbite are produced mechanically without the introduction of impurities, making them more effective in generating activation sites with varied coordination numbers and dangling bond characteristics for the photocatalytic reaction [72-74]. As shown in Fig. 11, experimental findings in this study confirm the HPT-treated catalyst presents a great generation of active radicals which are believed to be responsible for the degradation of plastics [68]. Analysis of PL further indicates that electron-hole recombination is suppressed by the formation of the high-pressure phase. In addition to the issues mentioned, the authors believe that the introduction of the high-pressure phase creates heterojunctions with anatase and rutile phases, which can improve the photo-induced charge separation for photoreforming [52,63].

Regarding the second question, this first study on the application of the high-pressure columbite phase for simultaneous H$_2$ production and plastic degradation confirms that the phase increases the catalytic activity compared to the initial anatase powder. The insertion of the high-pressure phase alters the band structure and increases the activities for H$_2$ production and PET plastic degradation without the addition of impurities, in contrast to doped catalysts [75]. To gain a deeper understanding of the level of performance of the HPT-treated catalyst with a high-pressure phase, its ability to generate $H_2$ is assessed by comparing it to previously reported data on $TiO_2$-based catalysts, as given in Table 2 [76-82]. There is a wide range of the performance of catalysts for $H_2$ photogeneration due to the differences in catalyst and catalytic process conditions. Table 2 confirms that the $TiO_2$ catalysts with the high-pressure phase show a high amount of $H_2$ production from photoreforming of PET. Compared to previous studies using a benchmark catalyst of P25 $TiO_2$ with the same co-catalyst of platinum, this HPT-processed $TiO_2$ showed a higher activity [77,78,82]. Among the data presented in Table 2, only P25 with gold co-catalyst shows better activity than the HPT-treated catalyst [76,77]. In summary, this study together with an earlier publication [83] introduces a new functionality for HPT-processed materials for the photoconversion of plastic waste, extending potential applications of severe plastic deformation processes to environmental issues [84].

Table 2. $H_2$ production from photoreforming on HPT-processed catalyst containing high-pressure $TiO_2$ phase, compared with reported data for $TiO_2$-based photocatalysts.

| Catalyst | $TiO_2$ Phases | Sacrificial Agent | Produced $H_2$ (mmol m$^{-2}$ h$^{-1}$) | Ref. |
|---|---|---|---|---|
| **Pt/P25** | Anatase, Rutile | Ethanol | 0.025 | [76] |
| **Au/P25** | Anatase, Rutile | Ethanol | 0.710 | [76] |
| **Au/P25** | Anatase, Rutile | Methanol | 0.294 | [77] |
| | | Ethanol | 0.213 | [77] |
| | | Ethylene Glycol | 0.456 | [77] |
| | | Glycerol | 0.609 | [77] |
| **Au/TiO$_2$** | Anatase | Methanol | 0.074 | [77] |
| | | Ethanol | 0.064 | [77] |
| | | Ethylene Glycol | 0.105 | [77] |
| | | Glycerol | 0.131 | [77] |
| **Au/TiO$_2$** | Rutile | Methanol | 0.031 | [77] |
| | | Ethanol | 0.015 | [77] |
| | | Ethylene Glycol | 0.062 | [77] |
| | | Glycerol | 0.109 | [77] |
| **Au/TiO$_2$** | Brookite | Methanol | 0.199 | [77] |
| | | Ethanol | 0.156 | [77] |
| | | Ethylene Glycol | 0.325 | [77] |
| | | Glycerol | 0.410 | [77] |
| **Pt/P25** | Anatase, Rutile | Methanol | 0.284 | [78] |
| **CuO$_x$/P25** | Anatase, Rutile | Methanol | 0.087 | [79] |
| **CuO$_x$/Mixed-TiO$_2$** | Anatase, Brookite | Methanol | 0.026 | [79] |
| **CuO$_x$/Mixed-TiO$_2$** | Anatase, Brookite, Rutile | Methanol | 0.017 | [79] |
| **CuO$_x$/Mixed-TiO$_2$** | Brookite, Rutile | Methanol | $0.08 \times 10^{-3}$ | [79] |
| **Pt/Mixed-TiO$_2$** | Anatase, Rutile | Cellulose | 0.002 | [80] |
| **Pt/Mixed-TiO$_2$** | Anatase, Rutile | Cellulose | 0.003 | [81] |
| **Pt/Anatase** | Anatase | Cellulose | 0.002 | [81] |
| **Pt/Anatase** | Anatase | Ethanol | 0.032 | [82] |
| **Pt/P25** | Anatase, Rutile | Ethanol | 0.040 | [82] |
| **Pt/Brookite** | Brookite | Ethanol | 0.060 | [82] |
| **Pt/Anatase** | Anatase | PET | 0.096 | This study |
| **Pt/HPT TiO$_2$** | Anatase, Columbite, Rutile | PET | 0.449 | This study |
| **Pt/HPT+Anneal TiO$_2$** | Anatase, Columbite, Rutile | PET | 0.214 | This study |

## 5. Conclusion

In this study, the effect of the high-pressure $TiO_2$ columbite phase containing oxygen vacancies on photoreforming for simultaneous $H_2$ production and PET plastic degradation was investigated for the first time. The introduction of the high-pressure phase altered the band structure, suppressed the recombination of electrons and holes, enhanced the reactive radical generation, and elevated the catalytic activity for both $H_2$ production and PET plastic degradation into various useful chemicals including formic acid, terephthalate, glycolic acid and acetic acid.

**Acknowledgments**

This study is supported partly by Mitsui Chemical, Inc. Japan, partly through Grants-in-Aid from the Japan Society for the Promotion of Science (JSPS), Japan (JP22K18737), and partly by the ASPIRE project of the Japan Science and Technology Agency (JST) (JPMJAP2332).